\documentclass[prx,twocolumn,pdflatex,showpacs,superscriptaddress]{revtex4}
\usepackage{amsmath, amsthm, amssymb,float}
\usepackage{amsfonts}
\usepackage{graphicx}
\usepackage{dcolumn}
\usepackage{bm}
\usepackage{textcomp}
\usepackage[normalem]{ulem}
\usepackage{mathrsfs}
\usepackage{color}

\usepackage{epsfig}
\usepackage{textcomp}
\usepackage{epstopdf}

\begin{document}

\title{Electromagnetic Solitons in Quantum Vacuum }

\author{S.  V. Bulanov}
\affiliation{Institute of Physics of the ASCR, ELI--Beamlines project, Na Slovance 2, 18221, Prague, Czech Republic}
\affiliation{National Institutes for Quantum and Radiological Science and Technology (QST),
Kansai Photon Science Institute, 8--1--7 Umemidai, Kizugawa, Kyoto 619--0215, Japan}
\affiliation{Prokhorov  General Physics Institute of the Russian Academy of Sciences,
Vavilov Str.~38, Moscow 119991, Russia}
\author{P. V. Sasorov}
\affiliation{Institute of Physics of the ASCR, ELI--Beamlines project, Na Slovance 2, 18221, Prague, Czech Republic}
\affiliation{Keldysh Institute of Applied Mathematics, Moscow, 125047, Russia}
\author{F. Pegoraro}
\affiliation{Enrico Fermi Department of Physics, University of Pisa, Italy
and National Research Council, National Institute of Optics, via G. Moruzzi 1, Pisa, Italy}
\author{H. Kadlecov\'{a}}
\affiliation{Institute of Physics of the ASCR, ELI--Beamlines project, Na Slovance 2, 18221, Prague, Czech Republic}
\author{S.\,S.\,Bulanov}
\affiliation{Lawrence Berkeley National Laboratory, Berkeley,CA 94720, United States of America}
\author{T. Zh. Esirkepov}
\affiliation{National Institutes for Quantum and Radiological Science and Technology (QST),
Kansai Photon Science Institute, 8--1--7 Umemidai, Kizugawa, Kyoto 619--0215, Japan}
\author{N. N. Rosanov}
\affiliation{Vavilov State Optical Institute, Kadetskaya liniya 5/2, 199053 Saint-Petersburg, Russia}
\affiliation{University ITMO, Kronverkskii prospect 49, 197101 Saint-Petersburg, Russia}
\affiliation{Ioffe Physical Technical Institute, Politekhnicheskaya ul. 26, 194021 Saint-Petersburg, Russia}
\author{G.\,Korn}
\affiliation{Institute of Physics of the ASCR, ELI--Beamlines project, Na Slovance 2, 18221, Prague, Czech Republic}

\date{\today}

\begin{abstract}
In the limit of extremely intense  electromagnetic fields  the Maxwell equations are modified due to the
photon-photon scattering  that makes  the vacuum refraction index depend  on the field amplitude.
In presence of  electromagnetic waves with small but finite wavenumbers
the vacuum behaves as a dispersive medium.
We show that the interplay  between the vacuum polarization and the nonlinear
effects in the interaction of counter-propagating electromagnetic waves
can result in the formation of Kadomtsev-Petviashvily solitons and, in  one-dimension
configuration, of  Korteveg-de-Vries type solitons that  can propagate over  a large distance without
changing  their shape.
\end{abstract}

\maketitle

\nopagebreak

\section{Introduction}

Fast progress in the laser and free electron laser technology aimed at developing sources of extremely high power
electromagnetic radiation has called into being a vast area of nonlinear physics related to the
behaviour of matter and vacuum irradiated by ultraintense electromagnetic fields~\cite{MTB06}.

Among the rich variety of
nonlinear effects induced by a relativistically strong light,
we will choose as the topic of the  present paper the formation and evolution of electromagnetic  solitary
waves in the quantum vacuum.

Relativistic electromagnetic solitons propagating in collisionless plasma have been
extensively studied 
theoretically~\cite{KLS, SRYT86, KAW, RNS-97, TZE, FLB-00, FB, FB-01, MLo-02, SPOO-02, FB-05, GL-06},
with computer simulations
~\cite{PFB-92, YS-99, TZE, NNM-01, TZE-02, TZE-04, 
JBK-04, SW-05, MA-06, YLIU-15, SVB-01, P-2019, S-2019, GSA-11a, GSA-11b, GSA-11c, VSA-13,  WU-19},
and in the experiments
on the laser-plasma interaction~\cite{MB-02, MK-07, ASP-07, LMC-07, MK-09, LR-10, GS-11}. Typically solitons in
relativistic plasmas can be regarded as  pulses of electromagnetic radiation trapped inside the cavities
formed in the plasma electron density by the pulse ponderomotive pressure. In the limit of small but finite
soliton amplitude they are described by the Nonlinear Schroedinger Equation (for the properties of
the NSE solitons see Refs. \cite{GBW-74, VEZ-84, DODD-85, ZKUFN}). Relativistic electromagnetic solitons
can provide one of the ways of  anomalous   absorption of the laser energy  by transforming it
into energy of fast particles and into energy of high and low frequency electromagnetic radiation.

The properties of  solitons that are formally similar to the NSE solitons were analyzed  theoretically 
in Refs.~\cite{SS-00, NNR-98}. We note that NSE solitons are predicted to be formed in the quantum vacuum. 
They correspond to  electromagnetic pulses trapped in the  local modulations of the refraction index of the vacuum.
In classical electrodynamics the vacuum refraction index equals unity, i.e.
 electromagnetic waves do not interact with
each other. On the contrary, in quantum electrodynamics (QED) electromagnetic waves interact in vacuum
via virtual electron-positron pair excitation  which is related to vacuum
polarization \cite{BLP, DG00, MSh06, PMHK12, KH16}.
In the other words, the electromagnetic field can excite a virtual electron-positron plasma.
The experiments on the detection of the photon-photon scattering using high power laser facilities
\cite{AAM-85, LBL-06, TH-06, TFM-08, MKSC11, SHS-16, BS-18, GKK18}
is  one of the most attracting goals in fundamental science.

The electromagnetic field intensity required for the  observation of the vacuum polarization is
characterized by the QED critical electric field. It is also known as the Schwinger
 field \cite{BLP} $E_S=m_e^2c^3/e\hbar$, where $e$ and $m_e$ are the electron charge and mass,
 $c$  the speed of light in vacuum, and $\hbar$  is the Planck constant. The
corresponding normalized wave amplitude $a_S=eE_S/m_e\omega c=m_ec^2/\hbar \omega$ and light intensity are
$5.1\times 10^5$ and  $10^{29}$ W/cm$^2$, respectively. By virtue of the Lorentz invariance,
a plane electromagnetic wave does not induce the vacuum polarization. In other words,
there is no self-action of a single plane wave. The situation becomes different for counter-propagating
electromagnetic pulses, when they mutually change the vacuum refraction index seen by the
other wave. The refraction index depends  nonlinearly  on the colliding electromagnetic
wave amplitude~\cite{BLP, DG00, BBBB70},
and the resulting  wave self-action can lead to  wave steepening and wave breaking~\cite{KKB19, KBK19}.
In the long-wavelength limit the QED vacuum is a dispersionless medium, i.e. the phase and group velocity of the
electromagnetic wave are equal. The vacuum dispersion effects seen at  small but finite photon
momentum have been analysed in
Refs.~\cite{MME81, GS99}. These effects  can also be found by using
an approach developed in Refs.~\cite{Na69, R70}. In general, the nonlinearity and dispersion balance provides
the condition for  the  formation of solitary waves~\cite{GBW-74, VEZ-84, DODD-85},
which can propagate over large distance without  changing their shape.
Along with the solitons described by the NSE equation,  were solitons described
by the Korteveg-de-Vries (KdV) equation~\cite{KdV85} (a generalization of the KdV equation to the
multidimensional case is known as the Kadomtsev-Petviashvili (KP) equation~\cite{KP}) \cite{GBW-74, VEZ-84,
DODD-85}.

Below we show that the vacuum polarization and  the nonlinear
effects in the interaction of counter-propagating electromagnetic waves
can result in the formation of the relativistic electromagnetic solitons and nonlinear waves
described by the KP,  KdV, and dispersionless Kadomtsev-Petviashvili (dKP)  equations.
Realizing conditions for the soliton formation in the super-strong laser beam collisions 
we might be able to understand better vacuum behavior testing the appearance 
of excitation of the electron positron Dirac sea.

The paper is organized as follows. In section II we discuss the EM wave dispersion in the QED vacuum as well as the long wavelength limit. 
The nonlinear EM waves in vacuum are discussed in section III. First, equations of nonlinear electrodynamcis are written down, 
then the case of the counter-propagating EM waves is investigated. In section IV we discuss EM waves in the QED vacuum described 
by  Kadomtsev-Petviashvili, dispersionless Kadomtsev-Petviashvili, and Korteveg-de-Vries equations. We conclude in section V.


\section{Electromagnetic wave dispersion in the QED vacuum }


\subsection{Dispersion equation}

The dispersion equation giving  the  relationship between the frequency $\omega$ and   the wave vector ${\bf k}$
of  a relatively high frequency small amplitude electromagnetic wave colliding in the QED vacuum with
 a low frequency wave can be written in the form
\begin{equation}
\label{eq:DE}
\omega^2-k^2 c^2-\frac{\mu_{\parallel, \perp}^2c^4}{\hbar^2}=0.
\end{equation}
In this case, the low frequency wave is approximated by the crossed field wave with
electric ${\bf E}$ and magnetic ${\bf B}$ fields orthogonal to each other of of equal amplitude, $E_0=B_0$. Here $\mu_{\parallel, \perp}$ is the ``invariant photon mass'' ~\cite{R70} (on the the effects 
of the field inhomegeity and the  limits of applicability of the crossed field 
approximation see Refs.~\cite{FK-15, DiP-18}).
The subscripts ${\parallel, \perp}$ of $\mu_{\parallel, \perp}$ correspond
to the parallel and perpendicular polarizations of the
colliding electromagnetic waves in the reference frame where they are counter-propagating to each other. For the sake of brevity we assume below that the wave
polarizations are parallel and denote by
 $\mu$ the invariant photon mass.

The invariant mass depends on the photon
frequency (it is the photon energy expressed in terms of the quantum parameter $\chi_\gamma$).
The invariant
\begin{equation}
\chi_\gamma=\frac{\hbar\sqrt{-k_\rho F^{\rho\sigma}F_{\sigma\tau}k^\tau}}{m_ecE_S},
\end{equation}
characterizes
the QED processes
of  photons interacting with an  electromagnetic field. Here $k^\rho$ is the 4-moment of the photon,
$F_{\rho \sigma}$ is the electromagnetic field tensor given by
\begin{equation}
\label{eq:Fmn}
F_{\rho \sigma}=\partial_{\rho} A_{\sigma}-\partial_{\sigma} A_{\rho},
\end{equation}
with $A_{\rho}$ being the 4-vector potential of the electromagnetic field,
 $\rho=0,1,2,3$, $\partial_{\rho}$ denotes partial derivative with respect
to the 4-coordinate $x_{\rho}$.  Here and below  summation over repeating indices is assumed.

For a  photon counter-propagating to the crossed $E_0$ - $B_0$ fields,  the invariant equals
\begin{equation}
\label{eq:ChiG}
\chi_\gamma=\frac{ E_0}{ E_S}\frac{\hbar (\omega+k_x c)}{m_e c^2},
\end{equation}
where $k_x$ is the $x$ component of the wave vector.

According to Ref.~\cite{R70} the square of the invariant photon mass $\mu$ is given by
\begin{equation}
\label{eq:Mu2}
\mu^2=\frac{\alpha m_e^2}{6\pi}\int_1^{\infty} du \frac{8u-2}{\zeta u\sqrt{u(u-1)}}\frac{df}{d\zeta}
\end{equation}
with
\begin{equation}
\label{eq:zuchi}
\zeta=\left(\frac{4 u}{\chi_{\gamma}}\right)^{2/3}
\end{equation}
and
\begin{equation}
\label{eq:fz}
f(\zeta)=i \int_0^{\infty} dt \exp \left[-i\left(\zeta t+ \frac{t^3}{3} \right)\right].
\end{equation}
In the r.h.s. of Eq. (\ref{eq:Mu2}) $\alpha=e^2/\hbar c\approx 1/137$ is the fine structure constant.

The function $f(\zeta)$ can be written as a linear combination of the Airy function $Ai(\zeta)$ and 
the inhimogeneous Airy function $Gi(\zeta)$, as
\begin{equation}
\label{eq:fzeta}
f(\zeta)=\pi\left[i Ai(\zeta)+Gi(\zeta) \right].
\end{equation}

Using  the  analytical properties of the Airy functions $Ai(\zeta)$ and 
 $Gi(\zeta)$ (see Refs.~\cite{AbSte-54, NIST} and Appendix \ref{AppA};  $Gi(\zeta)$ is also known as the Scorer 
 function~\cite{SCORER}) 
 we can present
the dependence of $f(\zeta)$ on the variable $\zeta$ in the limit
$\zeta \to +\infty$ (i.e. in the limit $\chi_{\gamma}\ll 1$) as
\begin{equation}
f(\zeta)=\frac{1}{\zeta}+\frac{2}{\zeta^4}
+ ...
+i\frac{\pi}{2 \zeta^{1/4}}\exp\left( -\frac{2}{3}\zeta^{3/2}\right).
\end{equation}
and its derivative is then
\begin{align}
f^{\prime}(\zeta)&=-\frac{1}{\zeta^2}-\frac{8}{\zeta^5}- ...
-i\frac{\pi}{8 \zeta^{5/4}}\exp\left( -\frac{2}{3}\zeta^{3/2}\right) \nonumber \\
-i&\frac{\pi}{2}\zeta^{1/4}\exp\left( -\frac{2}{3}\zeta^{3/2}\right).
\end{align}
We can neglect the last term in the derivative in the limit $\chi_{\gamma}\ll 1$.
Substituting this expression into  the integrand in the r.h.s. of Eq. (\ref{eq:Mu2}) and
calculating the integral we find expansions of the real  and the imaginary parts of the the  square of the   photon mass
at $\chi_{\gamma}\ll 1$,
\begin{equation}
\label{eq:Mu2Re}
\Re[\mu^2] = -\alpha m_e^2\frac{4}{45 \pi} \left[\chi_{\gamma}^2+\frac{1}{3}\chi_{\gamma}^4
+{\cal O}\left(\chi_{\gamma}^6\right)\right] \, ,
\end{equation}
\begin{equation}
\label{eq:Mu2Im}
\Im[\mu^2]=-\alpha m_e^2\frac{1}{8}\sqrt{\frac{3}{2}}\chi_{\gamma}\exp\left( -\frac{8}{3 \chi_{\gamma}}\right)
+\dots\, .
\end{equation}
Furthermore  we neglect the effects of  the exponentially small imaginary part (\ref{eq:Mu2Im})
which describes the electron-positron pair creation via the Breit-Wheeler process~\cite{BW34, NR70}.

We assume here that the
Poynting vector of the strong low frequency wave $c \, {\bf E}\times{\bf B}/4\pi$ 
is directed in the negative direction along the $x$-axis.
For definiteness we set $\mathbf{E}=\mathbf{e}_zE_0$, $\mathbf{B}=\mathbf{e}_yE_0$.
The high frequency electromagnetic wave propagates in the negative direction along the $x$-axis.

Substituting the real part of the photon mass given by the first two terms in Eq. (\ref{eq:Mu2Re}) 
into the dispersion equation (\ref{eq:DE})
we obtain for the relationship between the electromagnetic wave frequency and the wave-number
\begin{equation}
\begin{split}
\label{eq:DElin}
\omega^2-k^2+\frac{4 \alpha W_0^2}{45\pi}(\omega+k_x)^2&\\
+ \frac{4 \alpha W_0^4}{135\pi}(\omega+k_x)^4&=0,
\end{split}
\end{equation}
where $W_0=E_0/E_S$ is the electric field of the cross field electromagnetic wave normalized
on $E_S$, $k^2=k_x^2+k_{y}^2$, the wave number and frequency are normalized
 on $\lambdabar_C^{-1}=m_c/\hbar$ and $c \lambdabar_C^{-1}=m_c^2/\hbar$. 

We can rewrite the dispersion equation (\ref{eq:DElin}) as
\begin{equation}
\label{eq:DElin-Norm}
\omega^2-k_x^2+\kappa_1(\omega+k_x )^2+\kappa_2(\omega+k_x )^4=k_{y}^2.
\end{equation}
Here parameters $\kappa_1$ and $\kappa_2$ are given by
\begin{equation}
\label{eq:kappa12}
\kappa_1=4 \alpha W_0^2/45\pi \qquad {\rm and} \qquad \kappa_2= 4 \alpha W_0^4/135\pi.
\end{equation}
The parameters $\kappa_1$ and $\kappa_2$ give a measure of the vacuum polarization and vacuum dispersion
effects, respectively. The term $k_{y}^2$ in the r. h. s. of Eq. (\ref{eq:DElin-Norm}) describes
the electromagnetic wave diffraction. 

{We note 
that the expression for the 
electromagnetic wave dispersion in the QED vacuum given by last terms in the r.h.s. of 
Eqs. (\ref{eq:DElin}, \ref{eq:DElin-Norm}) being the consequence of the 6-photon mixing 
is different from the dependence found in~\cite{MME81} and used in Ref.~\cite{NNR-98} for describing 
the envelope solitons. Due to the symmetry of dispersion equation (\ref{eq:DE}) with the invariant photon mass given 
by relationships (\ref{eq:Mu2}-\ref{eq:fz}) the first nonvanishing dispersion term corresponds to the 6-photon mixing.}

Using the relationships between the frequency $\omega$ and wave-number $k$ and the  partial derivatives
with respect to time  and spatial   coordinates,
\begin{equation}
\label{eq:om-to-dt}
\omega \leftrightarrow -i\partial_t, \qquad  k_x\leftrightarrow i\partial_x, \qquad {\rm and} \qquad  k_{y} \leftrightarrow i \partial_y
\end{equation}
we obtain from Eq. (\ref{eq:DElin-Norm})
\begin{equation}
\label{eq:KP-lin}
\partial_-\left(\partial_+ a- \kappa_1\partial_- a - 2 \kappa_2\partial_{---}a \right)=-\frac{1}{2}\partial_{yy}a\, .
\end{equation}
with $\partial_-=\partial_{x^-}$,  $\partial_+=\partial_{x^+}$, and $\partial_{---}=\partial^3_-$. 
$\partial_\pm=\partial_{x^\pm}=(\partial/\partial x^\pm)$.

In Eq. (\ref{eq:KP-lin})  $a({x^-},{x^+},y)$ is the $z$ component of the 4 vector potential.
Here and below we use the so-called Dirac's
light cone coordinates ${x^-}$ and ${x^+}$ defined as (see  e.g. Ref.~\cite{LCD})
\begin{equation}
\label{eq:x+x-}
x^+=\frac{x+ct}{\sqrt{2}},\quad
x^-=\frac{x-ct}{\sqrt{2}},
\end{equation}
As well known, the coordinates $(x,t)$ in the laboratory frame of reference
are related to  the coordinates $(x',t')$ in the frame of reference moving with
the normalized velocity $\beta$ as
\begin{equation}
\label{eq:LorTr}
\begin {split}
x'=&x \cosh \eta -c t \sinh \eta, \\
t'=&t \cosh \eta -(x/c) \sinh \eta, 
\end{split}
\end{equation}
where
\begin{equation}
\eta ={\rm ln}\sqrt{\frac{1+\beta}{1-\beta}}.
\end{equation}
The Lorentz transform  of the light-cone variables, $x^{\prime +}$, $x^{\prime -}$, defined in Eq. (\ref{eq:x+x-}) is
\begin{equation}
\label{eq:LorTr+-}
\begin {split}
x^{\prime +}=\frac{x'+ct'}{\sqrt{2}}=e^{- \eta} \frac{x+ct}{\sqrt{2}}=e^{- \eta}x^+, &\\
x^{\prime -}=\frac{x'-ct'}{\sqrt{2}}=e^{ +\eta} \frac{x-ct}{\sqrt{2}}=e^{+ \eta}x^- .&
\end{split}
\end{equation}
As a result
\begin{equation}
\left(\partial_-\right)^\prime=e^{- \eta}\partial_-\, 
\quad {\rm and} \quad 
\left(\partial_+\right)^\prime=e^{+ \eta}\partial_+\,  ,
\end{equation}

Now we introduce the field variables $u$ and $w$ defined as
\begin{equation}
\label{fields}
u=\partial_{-}a \quad {\rm and} \quad w=\partial_{+}a.
\end{equation}
They are related to the electric, $e_z=-\partial_t a$  (along $z$), and magnetic, $b_y = - \partial_x a$ 
(along $y$), fields by the following relations
\begin{equation}
\label{fields2}
u =\frac{e_z-b_y}{\sqrt{2}}, \quad w  = -\frac{e_z+b_y}{\sqrt{2}}.
\end{equation}
The Lorentz transform of the fields $u$ and $w$  is
\begin{equation}
\begin{split}
\label{eq:fields-LT}
u^{\prime}& =\frac{e_z^{\prime}-b_y^{\prime}}{\sqrt{2}}=e^{- \eta}\frac{e_z-b_y}{\sqrt{2}}=e^{- \eta} u,\\
w^{\prime} &=-\frac{e_z^{\prime}+b_y^{\prime}}{\sqrt{2}}=-e^{+ \eta}\frac{e_z+b_y}{\sqrt{2}}=e^{+ \eta} w.
\end{split}
\end{equation}
The  field product $ uw = (b_y^2-e_z^2)/{2}$,
\begin{equation}
\label{eq:fields-uw}
u^{\prime} w^{\prime}=uw,
\end{equation}
is Lorentz invariant in the $(t,x)$-plane. It is proportional to the first Poincar\'e invariant $\mathfrak{F}$
of the Maxwell equations, which will be introduced below.
We note that $W_0$ transforms like $w$:
\begin{equation}
W_0^\prime=e^\eta W_0\, .
\end{equation}
%


\subsection{Dispersionless vacuum in the long wavelength limit}


\subsubsection{Counter-propagating electromagnetic waves}

Eq. (\ref{eq:DElin}) is obtained within the framework of the approximation, which assumes that the parameter
$\chi_{\gamma}$ is small. Neglecting 
 the dispersion and diffraction effects we
can write  Eq.  (\ref{eq:DElin-Norm}) as
\begin{equation}
\label{eq:DElin-nDD}
(\omega+k_x c)\left[\omega (1+\kappa_1)-k_x c(1-\kappa_1)\right]=0.
\end{equation}
Taking into accounts the relations given by Eqs. (\ref{eq:om-to-dt}) and (\ref{eq:x+x-}) this equation leads to
the wave equation
\begin{equation}
\label{eq:DElin-wave}
\partial_-\left(\partial_+ a- \kappa_1\partial_- a\right)=0,
\end{equation}
with the solution
\begin{equation}
\label{eq:DElin-wave-sol}
a({x^-},{x^+})=f({x^+})+g({x^-}+\kappa_1{x^+}).
\end{equation}
Functions $f(x)$ and $g(x)$ are determined by the initial conditions $a_0(x)$ and $\dot a_0(x)$.
A dot denotes a differentiation with respect to time.

Using the fact that $\kappa_1\ll 1$,  the solution (\ref{eq:DElin-wave-sol}) can be written in the following form
\begin{equation}
\label{eq:DElin-wave-sol-beta}
a(x,t)=f({x+ct})+g(x-v t),
\end{equation}
where $v=c(1-\kappa_1)/(1+\kappa_1)$.

The Cauchy problem is determined by the initial conditions
\begin{equation}
\begin{split}
\label{eq:Cauchy1}
a_0(x)=&f(x)+g(x),\\
\dot a_0(x)=&cf^{\prime}(x)-v g^{\prime}(x).
\end{split}
\end{equation}
A prime here and below denotes a differentiation with respect to the function argument.

Since $a_0^{\prime}(x)=f^{\prime}(x)+g^{\prime}(x)$ we can find that
\begin{equation}
\begin{split}
\label{eq:fg}
f(x) =\frac{v}{c+v}a_0(x)+\frac{1}{c+v}\int ^x\dot a_0(s) ds,&\\
g(x) =\frac{c}{c+v}a_0(x)-\frac{1}{c+v}\int ^x\dot a_0(s) ds.&
\end{split}
\end{equation}
Substituting these expressions into Eq. (\ref{eq:DElin-wave-sol-beta}) we obtain the solution to
the wave equation (\ref{eq:DElin-wave})
\begin{equation}
\label{eq:DElin-wave-solfin}
\begin{split}
a(x,t)=\frac{v\, a_0(x+ct)+c\,a_0(x-v t)}{c+v}&\\
-\frac{1}{c+v}\int^{x+t}_{x-v t} \dot a_0(s) ds.&
\end{split}
\end{equation}
In the case $v=c$ it becomes a standard d'Alembert formula.

\begin{figure} [ht]
\includegraphics[width=0.5\textwidth]{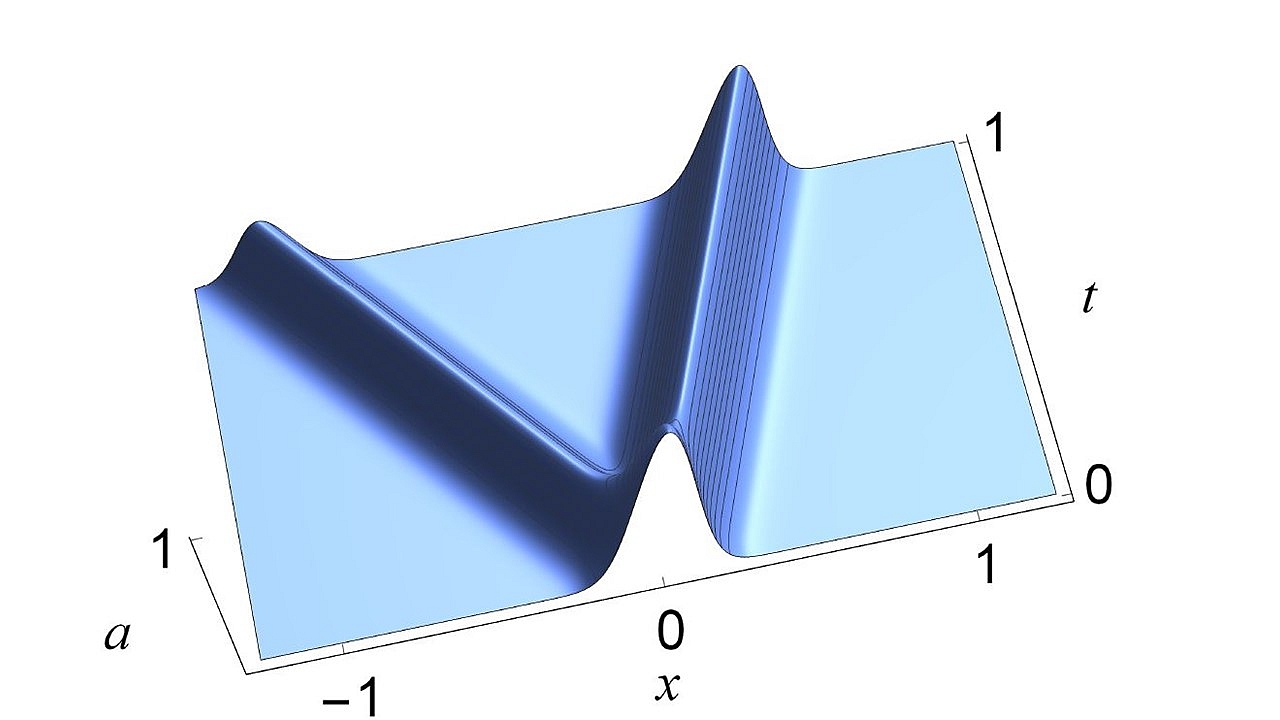}
\caption{Electromagnetic waves in the $(x,t)$ plane for $\dot a_0=0$ and $ a_0=\exp(-x^2/l^2)$ with $l=0.125$
and $\kappa_1=1/3$.
}
\label{Figure1}
\end{figure}

Fig. \ref{Figure1} shows the electromagnetic waves in the $(x,t)$ plane, which are determined by the initial
conditions $\dot a_0=0$ and $ a_0=\exp(-x^2/l^2)$ with $l=0.125$
and $\kappa_1=1/3$, i. e. $v=0.5c$. There are two waves. One of them propagates from the right to the left
with the speed of light in vacuum.  The other wave  propagates from the left to the right with speed equal to $v$.
The ratio of their amplitudes is equal to $v/c$. 

The vacuum polarization in the field of interacting electromagnetic waves
changes the electromagnetic wave propagation velocity making  Cherenkov radiation possible  in  vacuum \cite{D02, AJM18, SCCR-19}.

\subsubsection{Frederick's diagrams}

In the long-wavelength limit, when $k_x\to 0$ one can neglect the last term in the l.h.s. of Eq. (\ref{eq:DElin}), i.e. neglect the dispersion but
retaining the diffraction effects. Then the dispersion equation can be written as
\begin{equation}
\label{eq:DElin-ND}
\omega^2-k^2c^2+\kappa_1(\omega+k c \cos{\theta})^2=0,
\end{equation}
where $k=|{\bf k}|=\sqrt{k_x^2+k_{\perp}^2}$. Here we introduce the angle between the wave vector direction
and the $x$-axis equal to $\theta=\arccos (k_x/k)$ in the polar coordinate system.

The solution of Eq. (\ref{eq:DElin-ND}) gives  the wave frequency
 \begin{equation}
\label{eq:DElin-ND-om}
\omega=-kc\frac{\kappa_1\cos{\theta}+\sqrt{1+\kappa_1\sin^2{\theta}}}{1+\kappa_1}.
\end{equation}

This relationship yields the phase diagram representing the dependence of normalized phase velocity
$\beta_{\rm ph}=\omega/kc$ on the angle $\theta$,
\begin{equation}
\label{eq:DElin-ND-betaph}
\beta_{\rm ph}=-\frac{\kappa_1 \cos{\theta}+\sqrt{1+\kappa_1\sin^2{\theta}}}{1+\kappa_1}.
\end{equation}

Frederick's diagram (it is the polar diagram for group velocity of the wave, for details e. g. see~\cite{BBK}) can be obtained by calculating the
group velocity ${\bf v}_g=\partial \omega/\partial {\bf k}$. Taking into account that $\omega=k v_{\rm ph}$, where
 the phase velocity $v_{\rm ph}$ depend on the direction of ${\bf k}$ only we obtain
\begin{equation}
\label{eq:DElin-ND-vg}
{\bf v}_g=v_{\rm ph}\frac{\bf k}{k}+{v}_{\perp g}\frac{{\bf k}_{\perp}}{k}.
\end{equation}
Here the perpendicular to the wave vector component equals
${\bf v}_{\perp g}=k\partial v_{\rm ph}/\partial {\bf k}$, i. e.
$|{\bf v}_{\perp g} |= v_{\perp g}=(\partial \omega/\partial {\theta})/k$

\begin{figure*} [ht]
\includegraphics[width=0.8\textwidth]{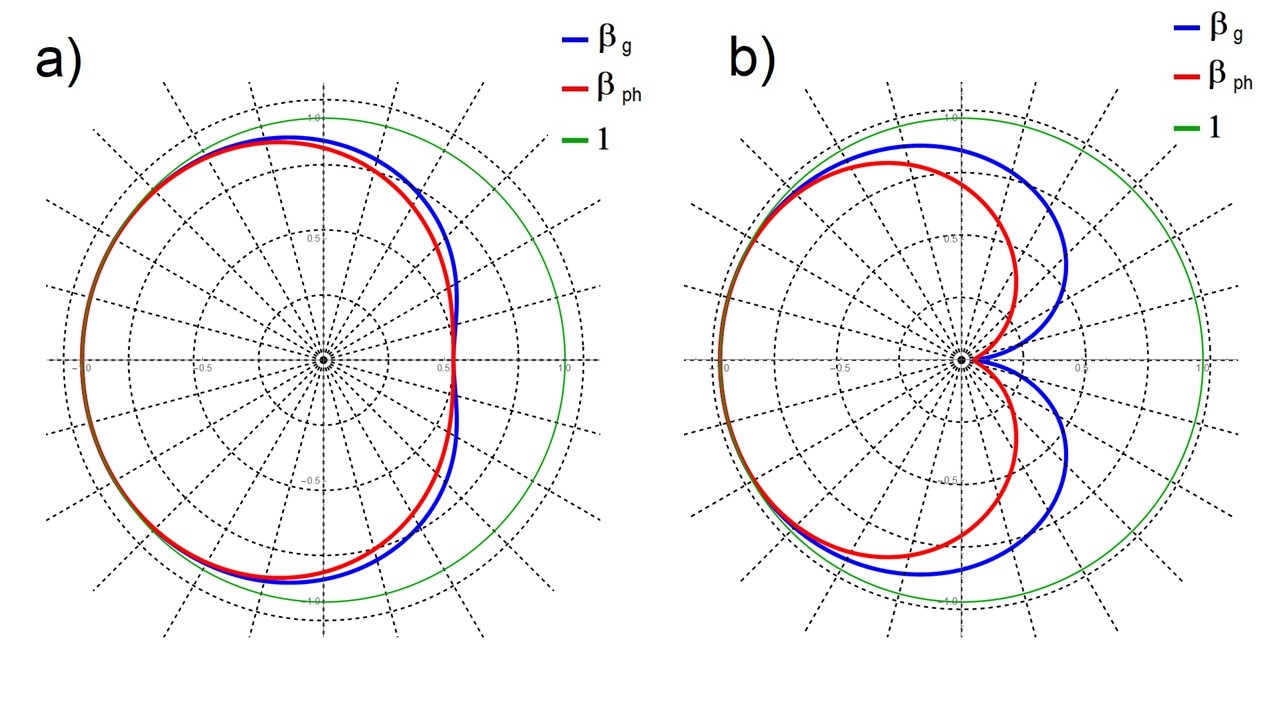}
\caption{Polar phase diagrams for group velocity $\beta_g$ (blue), phase velocity $\beta_{ph}$ (red)
and speed of light in vacuum $\beta=1$ (green): a)  $\kappa_1=0.3$; b)  $\kappa_1=0.9$.}
\label{Figure2}
\end{figure*}

For the normalized value of the perpendicular component $\beta_{g \perp}=v_{g \perp}/c$ we have
\begin{equation}
\label{eq:DElin-ND-betag-p}
\beta_{{\rm g} \perp}=\frac{\kappa_1 \sin{\theta}\left(\cos{\theta}+ \sqrt{1+\kappa_1\sin^2{\theta}}\right)}
{(1+\kappa_1) \sqrt{1+\kappa_1\sin^2{\theta}}}.
\end{equation}
Fig.~\ref{Figure2} presents the polar phase diagrams for the phase $\beta_{\rm ph}$ velocity and the  group velocity
$\beta_g=\sqrt{\beta_{\rm ph}^2+\beta_{\perp g}^2}$ : a) $\kappa_1=0.3$ and b)  $\kappa_1=0.9$.
As it is clearly seen the phase and group velocities are equal to each other for waves propagating along the
$x$-axis being equal to speed of light in vacuum for co-propagating waves, i.e. for
$\theta = 0 $ when $\beta_g=\beta_{ph}<1$ and $\beta_g=\beta_{ph}=-1$ for $\theta =\pi$.
In the cases $\theta \neq 0 $ and $\theta \neq \pi$, the phase velocity is smaller  than the group velocity.

\section{Nonlinear electromagnetic waves in vacuum}
\label{steepening}

\subsection{Equations of nonlinear electrodynamics}

\label{sec: HE-L}

Our consideration here is based on using the Euler--Heisenberg  Lagrangian describing
the electromagnetic field in the long-wavelength limit. It is given by \cite{H-E, BLP}
\begin{equation}
\mathcal{L}=\mathcal{L}_{0}+\mathcal{L}', \label{eq:Lagrangian}
\end{equation}
where
\begin{equation}
\mathcal{L}_{0}=-\frac{m_e^4}{16\pi\alpha}F_{\mu \nu}F^{\mu \nu}
\end{equation}
 is the  Lagrangian in classical electrodynamics, $F_{\mu \nu}$ is the electromagnetic field tensor determined
 by Eq. (\ref{eq:Fmn}).

In the Euler--Heisenberg theory, the QED radiation corrections are described by $\mathcal{L}'$ on the right hand side of Eq.(\ref{eq:Lagrangian}),
which can be written as \cite{BLP}
\begin{equation}
\begin{split}
\label{eq:HELagr}
\mathcal{L}^{\prime}=\frac{m_e^4} {8 \pi^2}{\cal M}( {\mathfrak e}, {\mathfrak b})=
\frac{m_e^4}{8\pi^2}\int^{\infty}_0 \frac{\exp{(-\eta)}}{\eta^3}\times \qquad&\\
\left[-(\eta {\mathfrak e}\cot \eta  {\mathfrak e}) (\eta {\mathfrak b}\coth \eta  {\mathfrak b}) +1-\frac{\eta^2}{3}({\mathfrak e}^2-{\mathfrak b}^2)\right] d \eta.&
\end{split}
\end{equation}
Here the invariant fields  $ {\mathfrak e}$ and $ {\mathfrak b}$
are expressed in terms  the Poincar\'e invariants
\begin{equation}
\label{eq:FGinv}
{\mathfrak F}=\frac{1}{4}F_{\mu \nu}F^{\mu \nu}, \,\,
{\mathfrak G}=\frac{1}{4}F_{\mu \nu}\tilde F^{\mu \nu}, \,\,
\tilde F^{\mu \nu}=\frac{1}{2}\varepsilon^{\mu \nu \rho \sigma}F_{\rho \sigma}
\end{equation}
 as
\begin{equation}
\label{eq:abinv}
{\mathfrak e}=\sqrt{\sqrt{{\mathfrak F}^2+{\mathfrak G}^2}-{\mathfrak F}} \,\,\, {\rm and}  \,\,\, {\mathfrak b}
=\sqrt{\sqrt{{\mathfrak F}^2+{\mathfrak G}^2}+{\mathfrak F}},
\end{equation}
respectively. 
Here  $\varepsilon^{\mu \nu \rho \sigma}$ is the Levi-Civita symbol in four dimensions.

Here and in the following text, we use the units $c=\hbar=1$, and the electromagnetic field
is normalized on the QED critical field $E_{S}$.

In the 3D notations the Poincar\'e invariants are 
\begin{equation}
\mathfrak{F}=\frac{1}{2}\left({\bf B}^2-{\bf E}^2\right),\quad
\mathfrak{G}={\bf B}\cdot{\bf E}.
\end{equation}

As explained in Ref.~\cite{BLP}
the  Euler--Heisenberg Lagrangian in the form given by Eq.~(\ref{eq:HELagr})
  should be used for obtaining an asymptotic
series over the invariant electric field
 $ {\mathfrak e}$ assuming its smallness.
The resulting expression is 
\begin{equation}
\label{eq:mathcalL}
\mathcal{L}^\prime=\kappa\left[\left( {\mathfrak F}^2
+\frac{ 7}{4} {\mathfrak G}^2\right) +\frac{8}{7} {\mathfrak F}
\left( {\mathfrak F}^2 +\frac{13}{16} {\mathfrak G}^2  \right) \right]+\dots
\end{equation}
with $\kappa= e^4/90 \pi^2 m_e^4$.

In the Lagrangian (\ref{eq:mathcalL}) the first two terms on the right hand side   and the last two correspond respectively to  four and to six photon mixing.


 \subsection{Counter-propagating electromagnetic waves}\label{CPEW}

In what follows we  consider the interaction of counter-propagating electromagnetic waves
with the same  linear polarization.  In this case the invariant ${\mathfrak G}$ vanishes identically.
This  field configuration  can be described  in a transverse gauge by a  vector potential having
a single  component, ${\bf A}=A {\bf e}_z$, with ${\bf e}_z$
the unit vector along the $z$ axis. In terms of the light cone coordinates (see  Eq. (\ref{eq:x+x-}))
the vector potential ${\bf A}$ is given by
\begin{equation}
\label{eq:Aa}
A=a(x^+,x^-).
\end{equation}
In these variables  the Lagrangian
(\ref{eq:Lagrangian}) takes the form
\begin{equation}
\mathcal{L}=-\frac{m^4}{4\pi\alpha}\left[wu-\epsilon_2 (wu)^2-\epsilon_3 (wu)^3  \right],
\label{eq:lightcone-Lagrangian}
\end{equation}
where the field variables $u$ and $w$ are defined by Eq. (\ref{fields}).
The dimensionless parameters
$\epsilon_2$ and $\epsilon_3$ in Eq.(\ref{eq:lightcone-Lagrangian}) are given by
\begin{equation}
\epsilon_2=\frac{2 e^2}{45\pi} =\frac{2}{45 \pi} \alpha \quad {\rm and} \quad
\epsilon_3=\frac{32 e^2}{315\pi} =\frac{32}{315 \pi} \alpha,
\end{equation}
 where $\alpha=e^2/\hbar c\approx 1/137$ is the fine structure constant,
i.e., $\epsilon_2 = 7\epsilon_3/8\approx 10^{-4}$.

The field equations can be found by varying the electromagnetic action
\begin{equation}
{\cal S}(a) = \int d x^+ \int  dx^-  \,{\cal L}(a),
\end{equation}
with respect to the vector potential $a(x^+,x^-)$ which gives
\begin{equation}
\label{EQMOT}
\partial_-(\partial_u \mathcal{L})+\partial_+(\partial_w \mathcal{L})=0.
\end{equation}
As a result, we obtain the system of equations
\begin{equation}
\label{eq:lightcone-1}
\partial_- w=\partial_+ u,
\end{equation}
\begin{equation}
\label{eq:lightcone-2}
\begin{split}
[1-uw(4\epsilon_2+9\epsilon_3uw)]\partial_+ u  =\qquad\qquad\qquad\qquad&\\
w^2(\epsilon_2 +3\epsilon_3 u w)\partial_- u +u^2(\epsilon_2+3\epsilon_3 u w)\partial_+ w,&
\end{split}
\end{equation}

Equation (\ref{eq:lightcone-1}),
is a consequence of the symmetry of the second derivatives,
$\partial_{-,+}a=\partial_{+,-}a$  and it expresses the vanishing
of the 4-divergence of the dual electromagnetic field tensor $\tilde F^{\mu \nu}$.

The solution to Eq. (\ref{eq:lightcone-1}) can be found to be
\begin{equation}
\label{eq:lightcone-3}
 w({x^+},{x^-})=\int^{x^-}\partial_+ u\, d {x^-}+w_0({x^+}),
\end{equation}
where $w_0({x^+})$ corresponds to the electromagnetic wave
propagating from the right to the left along the $x$-axis with a speed equal to the light speed in vacuum.


\subsection{The Hopf Equation}

The system of equations (\ref{eq:lightcone-1})-(\ref{eq:lightcone-2}) is a system of quasilinear equations. It admits a
rich variety of solutions including those solutions that  describe the formation of  singularities
during the electromagnetic field evolution (e.g. see Ref.~\cite{hodograph, hodograph20}). This system also admits
 solutions in the form of simple waves~\cite{KKB19}  in which  $w$ is a function of $u$, i.e. $w=w(u)$.
In this case, Eqs.~(\ref{eq:lightcone-1}) and~(\ref{eq:lightcone-2}) take the form
\begin{equation}
\label{eq:SW-1}
J \partial_- u=\partial_+ u,
\end{equation}
\begin{equation}
\label{eq:SW-2}
\partial_- u=\frac{1-uw(4\epsilon_2+9\epsilon_3uw)-Ju^2(\epsilon_2+3\epsilon_3 u w)}{w^2(\epsilon_2 +3\epsilon_3 u w)}\partial_+ u
\end{equation}
where $J=dw/du$ is the Jacobian.
Consistency of these equations implies that
\begin{equation}
\label{eq:Ja1}
u^2 J^2-\frac{1-u w \left(4\epsilon_2+9 \epsilon_3 uw\right)}{\epsilon_2+3 \epsilon_3 u w }J+w^2=0.
\end{equation}
Introducing the new variables
\begin{equation}
\label{eq:rl}
r=u w \quad {\rm and } \quad l=\ln \, u,
\end{equation}
for which 
\begin{equation}
J=\frac{1}{u^2}\left(\frac{dr}{dl}-r\right),
\end{equation}
  we can write the solution to Eq. (\ref{eq:Ja1})  as
\begin{equation}
\label{eq:Ja2}
\int^{uw}\frac{2\left(\epsilon_2+3 \epsilon_3r\right)dr}
{{\cal F}(r)}=l
\end{equation}
where
\begin{equation}
\label{eq:Ja3}
\begin{split}
{\cal F}(r)=1-2\epsilon_2r-3\epsilon_3 r^2 \pm \quad\quad\quad\quad\quad\quad\quad\quad\quad& \\
\sqrt{
1-8 \epsilon_2r+6(2 \epsilon_2^2 -3 \epsilon_3) r^2+48 \epsilon_2\epsilon_3 r^3 +45 \epsilon_3^2 r^4
},&
\end{split}
\end{equation}

Expanding this solution  up to linear  terms in $\epsilon_2$ and  $\epsilon_3$ we obtain for
the Jacobian $J$
\begin{equation}
\label{eq:Jacobi1}
J=w^2 (\epsilon_2+3\epsilon_3 uw)+\dots \, .
\end{equation}
We assume that the electromagnetic wave corresponding to the variable $u$ in Eq.~(\ref{eq:Jacobi1})
counter propagates with respect to the unperturbed wave $w_0= w_0(x^+)$, taken to depend on $x^+$ only.
Using equations~(\ref{eq:DElin-wave}) and~(\ref{eq:lightcone-3}) and  the smallness of the parameter $\epsilon_2$
we can obtain that
\begin{equation}
\label{eq:wxpxm}
w({x^+},{x^-})\approx w_0({x^+})+\epsilon_2 w_0({x^+})u({x^+}, {x^-}).
\end{equation}

Further we consider  the electromagnetic wave $w_0({x^+})$  to have a constant amplitude $w_0=-\sqrt{2}W_0=$~const,
i. e. to correspond to  crossed fields. In this case, the Jacobian~(\ref{eq:Jacobi1}) is
\begin{equation}
\label{eq:Jacobi2}
J=2\epsilon_2 W_0^2-6\sqrt{2}\epsilon_3 W_0^3\,u+\dots \, .
\end{equation}
 Substitution of $J$ given by Eq. (\ref{eq:Jacobi2}) into Eq. (\ref{eq:SW-1}) yields
\begin{equation}
\label{eq:SW-3}
\partial_+ u-\left(2\epsilon_2W_0^2- 6\sqrt{2}\epsilon_3 W_0^3 u\right)  \partial_- u=0.
\end{equation}
{In the third term describing nonlinear effects we retain the 6-photon mixing effects because as 
it is shown in Refs.~\cite{KKB19, KBK19} the 4-photon mixing photon effects is of higher order 
of a small parameter $\alpha$. We note that the envelope solitons considered in Refs.~\cite{NNR-98, SS-00} 
have been 
discussed within the framework of the 4-photon mixing approximation.}

Introducing the new dependent variable
\begin{equation}
\label{eq:baru}
\bar{u}=-\left(2\epsilon_2W_0^2- 6\sqrt{2} \epsilon_3W_0^3 u\right)
\end{equation}
we can rewrite Eq.~(\ref{eq:SW-3}) as  the Hopf equation
\begin{equation}
\label{eq:SW-5}
\partial_+ \bar{u} + \bar{u}\partial_-  \bar{u}=0.
\end{equation}

Eq.~(\ref{eq:SW-5}) has two groups of symmetries. It means that it remains the same if we make two sets of substitution:
\begin{equation}
\label{eq:Gal}
\bar{u}\to u_0+\bar{u}\, \qquad x^-\to  x^- + u_0x^+\, ;
\end{equation}
and
\begin{equation}
\label{eq:sc}
x^-\to \frac{x^-}{X^-}\, ,\quad x^+\to \frac{x^+}{X^+}\, ,\quad \bar{u}\to\frac{X^+}{X^-}\bar{u}\, ;
\end{equation}
where $u_0$, $X^-$ and $X^+$ are arbitrary real constants. 
Under the transformations (\ref{eq:Gal},\ref{eq:sc})  solutions of Eq.(\ref{eq:SW-5}) go into solutions.

 As is well known, the Hopf equation describes the  steepening of nonlinear waves (see Ref.~\cite{GBW-74}).
 In the case of a finite amplitude electromagnetic wave in the  QED vacuum this equation has been obtained
 and analysed in Refs.~\cite{KKB19, KBK19}.

The solutions to the Hopf equation (\ref{eq:SW-5}) can be obtained as follows~\cite{GBW-74, BBK}. The l.h.s. of the Hopf equation is a full derivative of the function 
 $\bar{u}(x^+,x^-)$
along the  characteristics of  Eq. (\ref{eq:SW-5}) determined by the equation
\begin{equation}
\label{eq:HOPF-1}
\frac{dx^-}{dx^+}=\bar{u}\, .
\end{equation}
The function $\bar{u}$ in Eq.~(\ref{eq:HOPF-1}) is constant, defined by initial conditions at $x^+=0$, and this equation can be rewritten as
\begin{equation}
\label{eq:HOPF-4}
\frac{dx^-}{dx^+}=\bar{u}(0,x_0^-)\equiv\bar{u}_0(x^-)\, .
\end{equation}
Relationship between variables $x^-$ and $x^+$ on the characteristics can be represented as
\begin{equation}
\label{eq:HOPF-2}
x^-=x_0^-+\bar{u}_0(x_0^-)x^+\, ,
\end{equation}
whereas general solution of the Hopf equation, defined by the initial condition $\bar{u}_0(x^-)$, 
can be written implicitly as
\begin{equation}
\label{eq:HOPF-3}
\bar{u}(x^-,x^+)=\bar{u}_0\left(x^--\bar{u}(x^-,x^+)x^+\right)\, .
\end{equation}

The value $x_0^-$, appearing in Eqs.~(\ref{eq:HOPF-2})-(\ref{eq:HOPF-4}), is the $x^-$-coordinate on 
the characteristic at $x^+=0$. In other words the variables $(x^+,x_0^-)$ are the  Lagrange coordinates. 
 Eq.~(\ref{eq:HOPF-2}) gives relationship between the Euler $(x^+,x^-)$ and Lagrange coordinates $(x^+,x^-_0)$.
 
Evaluating the gradient of the function $\bar{u}(x^-,x^+)$ we obtain
\begin{equation}
\label{eq:HOPF-4}
\partial_- \bar{u}=(\partial_- x_0^-)\, (\partial \bar{u}_0/\partial x_0^- ),
\end{equation}
 where
\begin{equation}
\label{eq:HOPF-5}
\partial_- x_0^-=\frac{1}{1+ x^+(\partial\bar{u}_0/\partial x_0^-)}
\end{equation}
 is the Jacobian of the transformation from the Lagrange to the Euler variables. In the region where
 $\partial \bar{u}_0/\partial x_0^-$ is negative the gradient (\ref{eq:HOPF-4}) grows.
 The growing of the Jacobian corresponds to the  steepening  of the wave  and to the generation of
high order harmonics (e.g. see discussion in Refs.~\cite{BBK, KKB19}).
 At the coordinate
\begin{equation}
{ x^+}_{br}=1/|\partial\bar{u}_0/\partial x_0^-|
\end{equation}
 the wave gradient tends to infinity: i.e. the wave breaks. This  corresponds to  the so called  gradient catastrophe.
 Using the relationship (\ref{eq:x+x-}) between the light cone coordinates $(x^-,x^+)$ and variables $(x,t)$
 we can find that the breaking time equals $t_{br}=1/c|\partial\bar{u}_0/\partial {x_0}|$. Here $x_0$ is the Lagrange coordinate if the Euler coordinates are $(x,t)$.

 In a dispersive medium the nonlinear wave steepening can be balanced by the dispersion effects resulting in formation of quasistationary nonlinear waves  such as collisionless shock waves and solitons.


 \section{Electromagnetic waves in the QED vacuum described by the Kadomtsev-Petviashvili, 
 the dispersionless
 Kadomtsev-Petviashvili,  and the Korteveg-de Vries  equations}
\label{KdV}

Below we show that combining the effects of difraction, dispersion and nonlinearity we obtain
  the nonlinear KP,  dKP,  and the KdV wave  equations.

  The Cauchy problem for these equations  can be solved
exactly~\cite{KdV67} (see also Refs.~\cite{GBW-74, DODD-85, VEZ-84} and the literature cited therein).
The solution includes in particular breaking nonlinear waves and  interacting solitons.

 From the  theory of nonlinear waves we know
 that in the limit of small but finite nonlinearity,  the  terms that describe  the dispersion and the  diffraction appear in the
 nonlinear wave equations additively. As far  as  it concerns the diffraction,  it can be implemented into the
 wave equation by considering the case when the electromagnetic potential (\ref{eq:Aa}) depends not only on the
 variables $x^+$ and $x^-$ but also on the transverse coordinate $y$ 
 (i.e. the beam size the $z$ direction is substantially larger than along the $y$ coordinate). 
 The nonlinear effects resulting in the finite
 amplitude wave breaking are described by Eq. (\ref{eq:SW-5}).

 \subsection{Kadomtsev-Petviashvili equation}
 \label{KPE}

 Combining Eqs. (\ref{eq:KP-lin}) and  (\ref{eq:SW-5})
 we obtain the  Kadomtsev-Petviashvili equation
\begin{equation}
\label{eq:KP}
\partial_-\left(\partial_+\bar{u}+\bar{u}\partial_- \bar{u} - \partial_{---} \bar{u}\right)=-\partial_{yy}\bar{u}.
 \end{equation}
In Eq.   (\ref{eq:KP})
the variables are normalized as $x^- \to x^-/L$, $x^+\to x^+/L$,  $y\to \sqrt{2}y/L$  with
$L=(2 \kappa_2)^{1/2}$. If one uses the units with $\hbar=c=1$, and fields are measured in $E_S$, then 
the coefficient $\kappa_2$ defined in Eq. (\ref{eq:kappa12}) equals
$\kappa_2=(4\alpha/135\pi) W_0^4m_e^{-2}$. 

Eq.~(\ref{eq:KP}) remains unchanged under the transform~(\ref{eq:Gal}). It is invariant also against the transforms:
\begin{equation}
\begin{split}
x^+\to x^+/X\, ,&\quad
x^-\to x^-/X^{1/3}\,,\\
y\to y/X^{2/3}\,,&\quad
\bar{u}\to \bar{u}/X^{2/3}\, ,
\end{split}
 \end{equation}
where $X$ is an arbitrary positive number.

In non normalized variables Eq. (\ref{eq:KP}) takes the form
\begin{equation}
\label{eq:KPnn}
\begin{split}
\partial_-\left[
\partial_+u-\left(\frac{4e^2}{45\pi}\,W_0^2-\frac{32 \sqrt{2}e^2}{105\pi}\,W_0^3\,u\right)\,\partial_-u \right.&\\
\left.
-\frac{8e^2}{135\pi m_e^2}\,W_0^4 \partial_{---} u
\right]=-\frac{1}{2}\partial_{yy}u&.
\end{split}
 \end{equation}

 Typical examples of the Kadomtsev-Petviashvili equation  solitons are discussed in Appendix \ref{AppB}.

To demonstrate the Lorentz-invariance of the problem under consideration it is convenient to rewrite Eq.~(\ref{eq:KPnn}) in the equivalent form
\begin{equation}
\label{eq:linv}
\begin{split}
&\partial_+u-\left(\frac{4e^2}{45\pi}\,W_0^2-\frac{32 \sqrt{2}e^2}{105\pi}\,W_0^3\,u\right)\,\partial_-u \\
&-\frac{8e^2}{135\pi m_e^2}\,W_0^4 \partial_{---} u
=-\frac{1}{2}\partial_{yy}a\,,
\end{split}
 \end{equation}
\begin{equation}
\partial_-a=u.
 \end{equation}
 As it can be seen from Eqs.(\ref{eq:LorTr+-}-\ref{eq:fields-LT})  each of the terms in the equation (\ref{eq:linv}) is Lorentz-invariant in the $(t,x)$-plane. 
 
Moreover, the way of derivation of this equation shows that, if we skip the nonlinear term, then remaining 
 equation is completely Lorentz-invariant in the $(t,x,y)$-hyperplane. Returning back to the whole equation, 
 we may say, that it remains the same under action of any weak rotation in $(x,y,z)$-hyperplane 
 and any weak Lorentz-transformation in $(t,y,z)$-hyperplane. 
 The ``weak rotation'' means neglecting the quadratic terms relative to the rotation angles and to the angle $\eta$ in
  Eqs. (\ref{eq:LorTr}). As far it concerns ``weak Lorentz-transformation'', according to Ref. \cite{LL-CTF} 
  it is sufficient to    satisfy the requirement that the components of one vector be small compared to those 
  of another in just one frame of reference; by virtue of relativistic invariance, the four-dimensional formulas 
  obtained on the basis of such an assumption will be valid in any other reference frame.
 
In regard with a relationship between the KdV and KP solitons discussed in the present paper and the solitons 
which can be obtained with NSE, here we briefly discuss the  evolution of a packet of quasi-monochromatic waves 
described by the 2D KP equation Eq.~(\ref{eq:KP-norm}). We assume that the carrier wave 
wavelength is short enough : $k_0\ell\gg 1$. In this case,  a 
 multi-scale expansion technique can be applied~\cite{GBW-74} for finding the solution 
 describing the wave packet evolution.  It can be easily shown,  the approach developed in Refs.~\cite{ZK86,Schn11}
  in the  present case results in obtaining the 3D version of  the Nonlinear Schroedinger Equation (NSE). 
From the NSE analysis, in this case, it follows that the wave packets are neither subject to  self focusing nor to bunching and, hence, during their evolution the NSE solitons are not formed.
 
\subsection{Dispersionless KP equation}

Neglecting the dispersion effects in Eq.   (\ref{eq:KP}) we obtain the
so-called dispersionless KP equation
\begin{equation}
\label{eq:dKP}
\partial_-\left(\partial_+\bar{u}+\bar{u}\partial_- \bar{u} \right)=-\partial_{yy}\bar{u},
 \end{equation}
The dispersionless KP equation describes the nonlinear wave breaking in  a non one-dimensional
configuration~\cite{dKP1, dKP2, dKP3, GRAVA1, GRAVA2}.

As noted above, formally the nonlinear wave steepening and breaking correspond to the growth of the field gradient
and to the appearance of the gradient catastrophe. This process can be demonstrated by analysing the self-similar solution of Eq. (\ref{eq:dKP}) of the form
 \begin{equation}
\label{eq:dKP-sol1}
\bar{u}(x^+,x^-,{\bf x}_{\perp})=g(x^+) x^- -\frac{\sigma}{2} k_y^2 y^2,
 \end{equation}
where $g$ and $k_y$ are the longitudinal and the  transverse inverse scale-lengths of the field $\bar{u}$, 
and $\sigma=\pm 1$. Substituting (\ref{eq:dKP-sol1})
into Eq. (\ref{eq:dKP}) we obtain an ordinary differential equation for the function $g(x^+)$,
\begin{equation}
\label{eq:dKP-sol2}
g^{\prime}+g^2=\sigma k_y^2,
 \end{equation}
where a prime denotes a differentiation with respect to the variable $x^+$.
Its solution reads
\begin{equation}
\label{eq:dKP-sol3}
g=-k_y \tan \left[k_y x^+ -\arctan \left(\frac{g_0}{k_y} \right)\right]
 \end{equation}
 if $\sigma=-1$ and
\begin{equation}
\label{eq:dKP-sol3s}
g=k_y \tanh \left[k_y x^+ -{\rm arctanh} \left(\frac{g_0}{k_y} \right)\right]
 \end{equation}
for $\sigma=+1$. Here $g_0$ is equal to $g|_{x^+=0}$.

 In the case $\sigma=-1$,  as
\begin{equation}
\label{eq:tto}
x^+\to \frac{1}{2 \,k_y}\left[ \pi+2\, {\rm arctan} \left(\frac{ g_0}{k_y}\right) \right]
\end{equation}
the gradient of $g$ tends to minus infinity.
 This corresponds to the wave breaking and to   the  formation
 of the shock wave like structure.


 \subsection{Korteveg-de Vries equation}

If one neglects the effects  of the transverse inhomogeneity by assuming
$\partial_{yy}\bar{u}=0$, then Eq.~(\ref{eq:KP})
reduces  to the Korteveg-de Vries equation~\cite{KdV85}
\begin{equation}
\label{eq:KvV}
\partial_+\bar{u}+\bar{u}\partial_- \bar{u} - \partial_{---} \bar{u}=0\,.
 \end{equation}
 It has the same symmetries as Eq.~(\ref{eq:KP}) in Sec.~\ref{KPE} . 
 
 We may not distinguish solutions of this equations related with each other by these symmetries.  Eq.~(\ref{eq:KvV}) rewritten in  physical variables looks as:
\begin{equation}
\label{eq:KvVnn}
\begin{split}
\partial_+u-\left(\frac{4e^2}{45\pi}\,W_0^2-\frac{32\sqrt{2}e^2}{105\pi}\,W_0^3\,u\right)\,\partial_-u \qquad&\\
=\frac{8e^2}{135\pi m_e^2}\,W_0^4 \partial_{---} u&.
\end{split}
 \end{equation}

Eq.~(\ref{eq:KvV}) has the well known single soliton solutions~\cite{GBW-74, DODD-85, VEZ-84, KdV85}. They can be presented in terms of Eq.~(\ref{eq:KvVnn}), when $u\to 0$ at $|x^-|\to\infty$, and $W_0 u_m>0$, as:
\begin{equation}
\label{eq:KdV-1}
 u=\frac{u_m}{\cosh^2\left[q(x^-+ v x^+)\right]}\,,
\end{equation}
This is a  constant shape localized nonlinear wave propagating with constant velocity $(1-v)/(1+v)$.
Its amplitude, $u_m$, and $v$,
are related to each other, as well as the soliton width, $q^{-1}$, as
\begin{equation}
\label{D300}
\begin{split}
v=&\frac{4e^2}{45\pi}W_0^2\left(1+\frac{8\sqrt{2}}{7}W_0\bar{u}_{m}\right),\\
q^2=&\frac{3\sqrt{2}}{7}m_e^2 \frac{\bar{u}_{m}}{W_0}.
\end{split}
\end{equation}

Evaluation of the soliton characteristic width, $\ell_s=1/q$, yields
\begin{equation}
\label{eq:ells}
\ell_s\approx 2 \frac{1}{m_e} \sqrt{\frac{W_0}{\bar{u}_{m}}}\left(=2\lambdabar_C\sqrt{\frac{E_0}{E_m}}\right),
\end{equation}
whereas the soliton formation length  is approximately equal to
\begin{equation}
\label{D330}
\begin{split}
\ell_f\approx&\frac{100}{e^2}\, \frac{W_0^{-4}}{m_e}\left(\frac{W_0}{\bar{u}_m}\right)^{3/2}\\
&\left(
=\frac{100}{\alpha}\lambdabar_C
\left(\frac{E_S}{E_0}\right)^4
\left(\frac{E_0}{E_m}\right)^{3/2}
\right).
\end{split}
\end{equation}
Here $E_0$ and $E_m$ are
the amplitudes of the counter-propagating waves  $\lambdabar_C=\hbar/m_e c$ is the Compton wavelength.
Assuming $E_0/E_m=100$ and $E_0/E_S\approx 1$ we obtain
for the soliton width $\ell_s=1.5\times10^{-2}$ nm and for the soliton
formation length $\ell_f =4\mu$m. 

We note that the   field invariant $\mathfrak{F}$ for this soliton, as  determined by Eq. (\ref{eq:FGinv}), 
is negative  i. e. 
it can be considered as an electromagnetic object where the electron-positron pair creation 
can occur via the Schwinger mechanism~\cite{H-E, BLP, JSCHW, VSP}.

\section{Conclusions}
\label{Con}

We have obtained an analytical description of  relativistic electromagnetic solitons that can be formed in a
configuration consisting of two counter-crossing electromagnetic waves propagating in the QED vacuum.
These extreme high intensity electromagnetic waves in the QED vacuum are described by
partial differential equations that belong to the family of the canonical equations in the theory
of nonlinear waves such as the Hopf,  the Korteweg-de Vries, the dispersionless  Kadomtsev-Petviashvili,
and  the Kadomtsev-Petviashvili equations. 

{In the case of the soliton solution of the KdV and KP equations the 
nonlinearity effects are balanced by the wave dispersion. The description of the nonlinearity leading to the wave 
steepening requires to take into account the 6-photon mixing process 
(for details see Refs.~\cite{KKB19, KBK19}) within the framework of the theoretical model 
based on the Heisenberg-Euler Lagrangian (\ref{eq:HELagr}), which being principally dispersionless corresponds 
to the longwavelength limit.  An adequate approach for calculating the QED vacuum dispersion 
is based on the perturbation theory developed in Refs.~\cite{Na69, R70} where the expression for the invariant 
photon mass (\ref{eq:Mu2}), which is a pole of the photon Green's function in a crossed field, has been obtained. 
The approach elaborated in Refs.~\cite{Na69, R70} is valid as long as $\alpha \chi_{\gamma}^{2/3}\ll 1$. As known, e.g. see \cite{KH16} in the small amplitude long-wavelength limit, when one can neglect the effects  dispersion and the nonlinearity  is weak the approches based on the perturbation theory and on the Heisenberg-Euler paradigm are equivalent. Analysis of analytical properties of the $f(\zeta)$ function (see Appendix \ref{AppA} allowed us to derive the dispersion term in Eq. (\ref{eq:Mu2Re}) which leads to the wave equation in the form (\ref{eq:KP-lin}). The dispersion term combination with the nonlinear term results in the KdV and KdP equations. }

These equations have  a wide range of applications in mathematics and physics that spans  from fluid mechanics
to solid state physics and to plasma physics.
The soliton theory is also used in  quantum field theory~\cite{LRaider-96, VRubakov, string1, dvali-15}.
In the present paper we extend the field of applications of the KdV, KP and dKP equations  to  the QED vacuum.

{The QED vacuum polarization effects are planned to be studied with the next generation lasers 
(see for details \cite{PMHK12, KH16, TH-06, MKSC11, FK-16, BS-18, UMi-19}).

In particular these effects can be revealed by measuring the phase difference between the phase of the
electromagnetic pulse colliding with the counterpropagating
wave and the phase of the pulse which does not interact
with high intensity wave, as well as by analyzing the wave frequency spectrum with specific features corresponding to 
the soliton formation. }

 Revealing the change in the parameters of colliding extremely intense laser beams will shed 
 a light on the space-time properties and 
vacuum texture.

\bigskip
\begin{acknowledgments}

The work is supported by the project High Field Initiative (CZ.02.1.01/0.0/0.0/15\_003/0000449)
from the European Regional Development Fund, by the Program 
of Russian Academy of Sciences ``Mathematics and Nonlinear Dynamics''. H. K. was supported by the fellowship (award)
Czech edition of L'Or\'eal UNESCO For Women In Science 2019. SSB acknowledges support from 
the Office of Science of the U.S. DOE under Contract No. DE-AC02-05CH11231.

\end{acknowledgments}
\appendix


\section{Analytical properties of the $f(\zeta)$ function}
\label{AppA}

According to Eqs. (\ref{eq:fz}) and (\ref{eq:fzeta})  the function $f(\zeta)$ can be presented 
in terms of the Airy functions $Ai(\zeta)$ and $Gi(\zeta)$ (see also Ref. \cite{R72}). 
Integral representations of the standard Airy 
function $Ai(\zeta)$  and of the inhomogeneous Airy function $Gi(\zeta)$ (it is also known as the Scorer function) 
 are~\cite{AbSte-54, NIST} 
\begin{equation}
\label{eq:Ai}
Ai(\zeta)=\frac{1}{\pi}\int_0^{\infty} dt \cos\left(\zeta t+ \frac{t^3}{3} \right)
\end{equation}
and 
\begin{equation}
\label{eq:Gi}
Gi(\zeta)=\frac{1}{\pi}\int_0^{\infty} dt \sin\left(\zeta t+ \frac{t^3}{3} \right),
\end{equation}
respectively. They obey the differential equations 
\begin{equation}
\label{eq:Ai-eq}
Ai''-\zeta Ai=0
\end{equation}
and 
\begin{equation}
\label{eq:Gi-eq}
Gi''-\zeta Gi=-\frac{1}{\pi}.
\end{equation}
Here a prime denotes a differentiation with respect to the variable $\zeta$. The equations should be 
solved with the initial conditions corresponding to expressions (\ref{eq:Ai}) and (\ref{eq:Gi}) and to (\ref{eq:Ai-Tay}) and (\ref{eq:Gi-Tay}) below.

The functions $Ai(\zeta)$ and $Gi(\zeta)$ can be expanded into the Maclaurin series as follows. 

\begin{equation}
\label{eq:Ai-Tay}
Ai(\zeta)=\frac{3^{-2/3}}{\pi} 
\sum_{n=0}^{\infty} \Gamma\left(\frac{n+1}{3}\right)\sin\left(\frac{3 n-1}{3}\pi\right)\frac{(3^{1/3}\zeta)^n}{n !}.
\end{equation}
and 
\begin{equation}
\label{eq:Gi-Tay}
Gi(\zeta)=\frac{3^{-2/3}}{\pi} 
\sum_{n=0}^{\infty} \Gamma\left(\frac{n+1}{3}\right)\cos\left(\frac{3 n-1}{3}\pi\right)\frac{(3^{1/3}\zeta)^n}{n !}.
\end{equation}
In the limit $\zeta \to 0$, i.e. for $\chi_{\gamma} \to \infty$ with the relationship between $\zeta$ and $\chi_{\gamma}$ 
given by Eq. (\ref{eq:zuchi}) expressions (\ref{eq:Ai-Tay}, \ref{eq:Gi-Tay}) give
\begin{equation}
\begin{split}
f(\zeta)=&i\frac{3^{-2/3}}{2} \left[ \Gamma\left(\frac{1}{3}\right)\left(\sqrt{3}+i\right)+\right .\\ 
&\left . 
 \Gamma\left(\frac{2}{3}\right)\left(-\sqrt{3}+i\right)3^{1/3}\zeta\right] + \dots \, .
\end{split}
\end{equation}
For large $\zeta$, when $\zeta \to \infty$ and $|\arg\zeta|<\pi$, asymptotic expansions of  $Ai(\zeta)$ and $Gi(\zeta)$
yield
\begin{equation}
\label{eq:Ai-inf}
\begin{split}
 Ai(\zeta)=&\frac{\zeta^{-1/4}} {2 \pi} \exp\left( - \frac{2}{3}\zeta^{3/2}\right) \\ 
& \times \sum_{n=0}^{\infty}(-1)^n\Gamma\left(3 n+\frac{1}{2}\right)
 \frac{\left(9 \zeta^{3/2} \right)^{-n}}{(2 n)!},
\end{split}
\end{equation}
\begin{equation}
\label{eq:Gi-inf}
 Gi(\zeta)=\frac{1}{ \pi \zeta} \sum_{n=0}^{\infty}\frac{(3 n)!}{n!} \left(3 \zeta^{3} \right)^{-n}.
\end{equation}

As a result we obtain for the asymptotic expantion of the function $f(\zeta)$ in the limit $\zeta \gg 1$
\begin{equation}
f(\zeta)=\frac{1}{\zeta}+\frac{2}{\zeta^4}
+\frac{120}{\zeta^7}
+ ...
+\frac{i\pi}{2 \zeta^{1/4}}\exp\left( -\frac{2 \zeta^{3/2}}{3}\right).
\end{equation}
%

%

%
\begin{figure*} [h]
\includegraphics[width=0.9\textwidth]{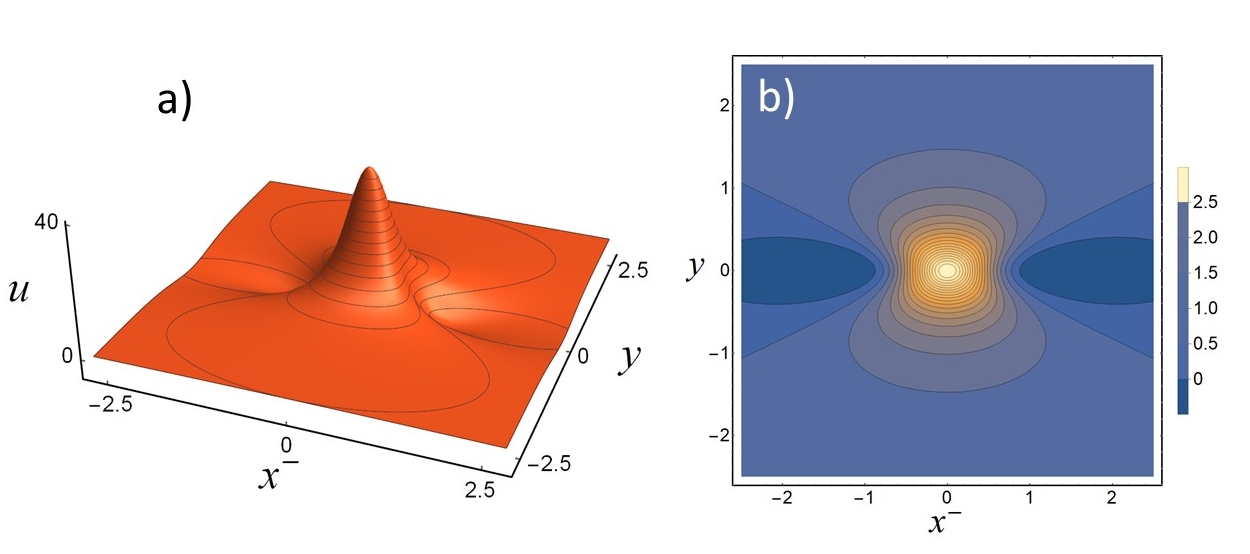}
\caption{Lump soliton  for $v=5$: a) $\bar{u}(x,y,0)$; b) contours of equal value of  $\bar{u}(x,y,0)$.
}
\label{FigureLump}
\end{figure*}
\begin{figure*} [h]
\includegraphics[width=0.85\textwidth]{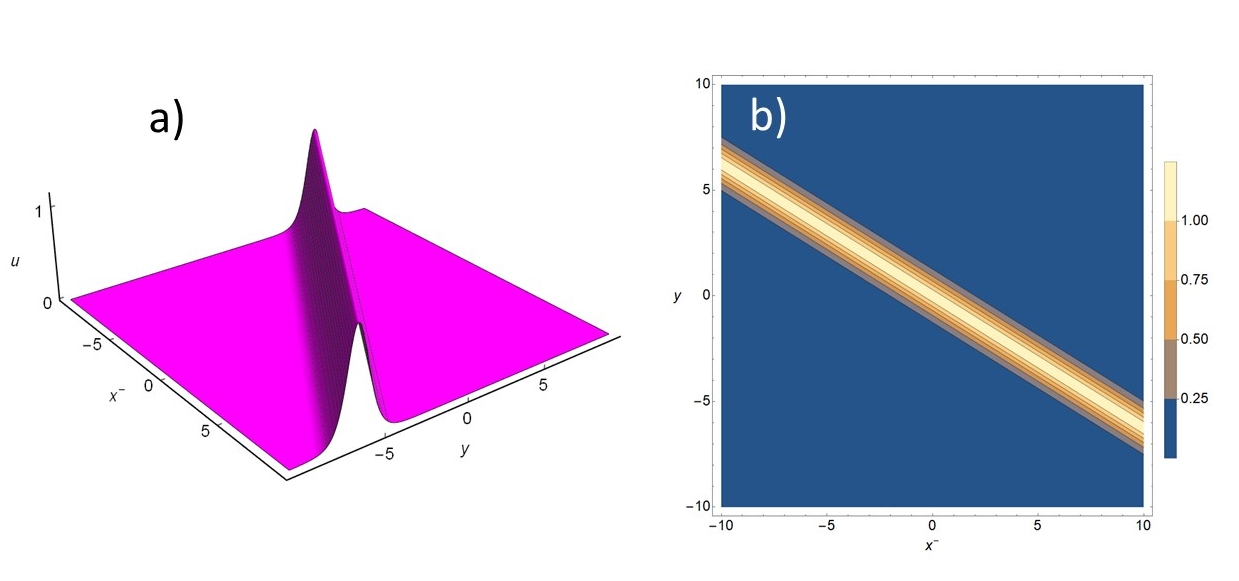}
\caption{Single soliton for KP equation}
\label{Figure4}
\end{figure*}
\begin{figure*} [h]
\includegraphics[width=0.85\textwidth]{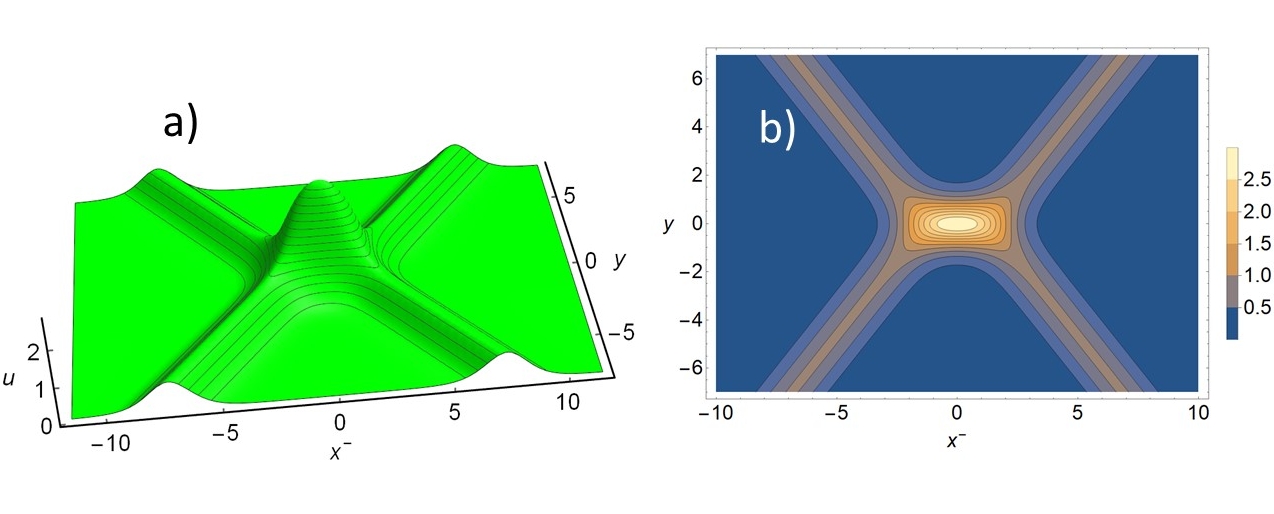}
\caption{Double soliton for KP equation}
\label{Figure5}
\end{figure*}

\section{Solitons of Kadomtsev-Petviashvili equation}
\label{AppB}

Here several typical examples of the solitons of Kadomtsev-Petviashvili
equation~\cite{DENG-05, MA-15, GBDEP-08}
are presented.
By rescaling independent and depended variables the KP equation can be reduced to
the normalized form
\begin{equation}
\label{eq:KP-norm}
\partial_-(\partial_+ u+6 u \partial_-+\partial_{---}u)=3 \partial_{yy}u.
\end{equation}

A rich variety of the soliton solutions of the KP equation
can be found with using the Hirota method~\cite{Hirota}, Backlund transformation or
the Wronskian technique~\cite{FREE-83}.

The localized solution of the KP equation (\ref{eq:KP-norm}) 
 is known as the ``lump'' and has the form~\cite{MANAKOV77, MA-15, ABL78, SAT79}
\begin{equation}
\label{eq:KP-lump}
u(x^+,x^-,y)=24 v \frac{3-v\left[(x{^-}+v x^+)^2-v y^2\right]}
{
\left\{
 3 +v\left[(x{^-}+v x^+)^2+v y^2 \right]
 \right\}^2
}.
\end{equation}
It is shown in Fig. \ref{FigureLump} a).
The propagation velocity in the $x^\pm$ variables $v$
of the    lump soliton and its maximum  amplitude are related   as $\bar{u}_m=8 v$.
The lump width is $\sqrt{3/v}$  and is inversely proportional to the square root of its amplitude as in the case
described by Eq. (\ref{eq:ells}).
At $x^-=vx^+$, along the $y$ axis the function $\bar{u}$ monotonically decreases as $\bar{u}\sim y^{-2}$.
In the plane $(x^-,y)$ it changes  sign on the hyperbola given by equation
\begin{equation}
\label{eq:hyp}
(x^- +vx^+)^2-v y^2=3/v.
\end{equation}
This hyperbola is clearly seen in Fig b), where the isocontours  of $\bar{u}(0,x^-,y)$ are plotted.

Multi-solitons for the KP equation can be found from equation $u=2\partial_{--}(\ln f)$ 
(see Ref. \cite{Hirota} and literature cited therein).
To single-soliton solution to the KP equation the function $f(x^-,x^+,y)$ equals
\begin{equation}
\label{eq:f}
f=1+\exp(\theta_1),
\end{equation}
where $\theta_1=k_1(x^-+\omega_1 x^+ +p_1 y)+\xi_1$ with $\omega_1=k_1^3+3p_1^2$.
Corresponding KP soliton is given by
\begin{equation}
u=\frac{k_1^2}{2 \cosh^2( \theta_1/2)}.
\end{equation}
It describes ``oblique KdV'' soliton whose maximum is localized on the line $\theta_1=0$.
It is shown in Fig.~\ref{Figure4}.

Double soliton solution for the KP equation is given by the function $f$ equal to
\begin{equation}
f=1+b_1\exp(\theta_1)+b_2\exp(\theta_2)+b_{12}\exp(\theta_1+\theta_2),
\end{equation}
where $\theta_i=k_i(x^-+\omega_i x^+ +p_i y)+\xi_i$ with $\omega_i=k_i^3+3p_i^2$ ($i=1,2$) and
\begin{equation}
\begin{split}
&b_{1}=-\frac{k_1+k_2+p_1-p_2}{k_1-k_2-p_1+p_2}, \quad  b_{2}=\frac{k_1+k_2-p_1+p_2}{k_1-k_2-p_1+p_2},\\
&b_{3}=-\frac{k_1-k_2+p_1-p_2}{k_1-k_2-p_1+p_2}.
\end{split}
\end{equation}
Double soliton for the KP equation is shown in Fig.~\ref{Figure5}.
\clearpage

\end{document}